\documentstyle[aaspp]{article}
\input{psfig}
\tighten
\begin{document}
\def\etal{{\it et al.\/}}
\def\cf{{\it cf.\/}}
\def\ie{{\it i.e.\/}}
\def\eg{{\it e.g.\/}}

\title{On the energy of neutrinos from gamma ray bursts}
\author{{\bf Mario Vietri}}
\affil{Universit\`a di Roma 3, Via della Vasca Navale 84, 00147 Roma, Italy \\ 
E-mail: vietri@corelli.fis.uniroma3.it }

\begin{abstract}
Ultra high energy protons accelerated at the shocks causing gamma ray bursts 
photoproduce pions, and then neutrinos {\it in situ}. I consider here the 
sources of losses in this process, namely adiabatic and synchrotron losses
by both pions and muons. When the shocks under consideration are external, 
{\it i.e.}, those between the ejecta and the surrounding interstellar medium,
I show that neutrinos produced by pion decay are unaffected by losses; those 
produced by muon decay, in the strongly beamed emission required by afterglow 
observations of GRB 971214, are limited in energy, but still exceed $10^{19}\; 
eV$. In particular, this means that ultra high energy neutrinos will be 
produced through afterglows.
\end{abstract}


\section{Introduction}

The discovery of the afterglows of Gamma Ray Bursts (GRBs; Costa \etal, 1997;
van Paradijs \etal, 1997; Frail \etal, 1997) and the disappearance of 
flares in their radio flux (Goodman 1997) have provided strong evidence in
favor of the fireball model (Rees and M\'esz\'aros 1992), whereby mass is
ejected from an as yet unknown source with large Lorenz factors, $\gamma 
\approx 100$. Several authors have pointed out that in the extremely
energetic events giving rise to these hyperrelativistic flows, high energy
neutrinos are likely to be produced. A first likely source is the merger of
the two neutron stars which may (Narayan, Paczynski and Piran 1992) give
rise to the bursts themselves, as originally suggested by Eichler \etal\/
(1989). Although insufficient to power the burst itself (Janka and Ruffert
1996), this may yet provide a copious source of low energy neutrinos, 
$\approx 1\; MeV$. Another source of neutrinos comes from $p-p$ collisions
inside the shocks which give rise to the observed burst proper (Paczynski
and Xu 1994); although it seems unlikely that these collisions may give
rise to the burst itself, yet they may originate a flux of higher energy 
neutrinos, $\approx 30\; GeV$, which is however currently unobservable
(Ostrowski and Zdziarski 1995).

A third, distinct source of high energy neutrinos exists.
It has recently been pointed out that gamma ray bursts may be responsible for
the acceleration of the highest energy cosmic rays observed so far (Vietri 1995,
Waxman 1995). It has been suggested (Waxman and Bahcall 1997) that these UHECRs 
may produce neutrinos through pion photoproduction, \ie, through the reaction
\begin{eqnarray}
p + \gamma & \rightarrow & n + \pi^+ \nonumber \\
\pi^+ & \rightarrow & \mu^+ + \nu_\mu  \\
\mu^+ & \rightarrow & e^+ + \nu_e + \bar{\nu}_\mu\;, \nonumber
\end{eqnarray}
inside the burst itself, \ie, against the $\gamma$--ray photons of the burst.
While the probability of this happening is not large (Vietri 1995), yet
Waxman and Bahcall (1997) pointed out that significant amounts of neutrinos with
energy $\approx 10^{14}\; eV$ ought to be produced, and might be revealed by 
detectors of the AMANDA class. This detection is however made difficult by the
fact that the Earth becomes opaque for this neutrino energy, so that the
technique employing upward--moving muons cannot be used. But, 
through the same mechanism, and simultaneously 
with these lower energy brethrens, neutrinos in the energy range $\approx
10^{19}\; eV$ (UHENs = Ultra High Energy Neutrinos) ought to be produced;
in Vietri (1998, paper I from now on) I showed that a significant fraction 
($f_\pi \approx 0.01$) of all energy initially released in UHECRs ought to 
be lost to neutrinos emitted during the burst proper, or during the first week
of the burst afterglow. The interesting point of this computation is that
it appears that the next generation of satellite--borne detectors of large
showers, such as AIRWATCH (Linsley 1997) may cover an area large enough
to allow detection of UHENs within the first year or so of operation. 

Shortly before paper I was accepted, Rachen and M\'esz\'aros (1998) have 
presented a detailed analysis of losses limiting neutrinos' energies, 
in the {\it internal} shock scenario for GRBs, which applies for instance
to the model of Waxman and Bahcall (1997). They conclude that these losses 
severely limit both the energies of individual neutrinos, and the total energy 
released by gamma ray bursts. Such losses arise 
because both the photoproduced pions, and the muons generated by the pion 
decay, before decaying suffer adiabatic and synchrotron losses, 
thusly imparting to neutrinos less than the energy they started out with. 
Given the short rest lifetimes of pions and muons, one maybe inclined to 
think these losses at all negligible, but, because of relativistic 
time--dilation, pions and muons moving with a Lorenz factor typical of 
ultra high energy cosmic rays ($\approx 10^{10}$) may survive more than 
$10^4\; s$, and cover sizeable distances ($\approx 10^{14}\; cm$) in the 
meantime. 

It is the purpose of this {\it Letter} to carry out the analysis 
of losses for {\it external} shocks, both during the burst proper and in the 
afterglow phase. It will be concluded that in the relatively tame 
environments generated by external shocks, neutrinos produced by pion decay 
are unaffected by losses, while neutrinos produced by muon decay are 
limited in energy by these processes, but still manage to exceed $10^{19}\; 
eV$. The major departure from the discussion in paper I is due to the 
sensational discovery (Kulkarni \etal, 1998) of the redshift $z = 3.43$ for the 
burst GRB 971214. This discovery forces us to choose different scaling values
for bursts, for the following reason. For cosmological parameters $\Omega = 0$ 
and $H_0 = 50\; km\; s^{-1}\; Mpc^{-1}$, the luminosity distance to this 
redshift is
$6.2\times10^{28}\; cm$; for $\Omega =1$ it would be $8.4\times 10^{28}\;
cm$. The burst as observed by the BeppoSAX GRBM had a peak luminosity of
$6\times 10^{-7}\; erg\; s^{-1}\; cm^{-2}$, and a total duration of $\approx
30\; s$ (Heise \etal, 1997), corresponding to a fluence of $\approx 10^{-5}
\; erg\; cm^{-2}$. With the lower, above--stated luminosity distance 
this converts to a total energy release of $4\times 10^{53}\; erg$, roughly 
the binding energy of two merging neutron stars. While strictly speaking still 
not an impossibility, these values seem to imply some beaming, a natural 
consequence of many GRBs' models. In fact, assuming a beaming factor $\theta 
\approx 0.1$, the total energy release becomes a more mundane $10^{51}\; erg$. 
Similar, although less extreme values have been derived, less cogently, for the 
burst GRB 970508 (Panaitescu, M\'esz\'aros and Rees 1998). These values 
will be adopted here. 

\section{On the acceleration of protons in fireballs}

Before considering losses, I need to establish some properties of the 
environment, within the fireball theory of gamma ray bursts. The maximum
proton energy, in the observer frame, was shown to be (Vietri 1995)
\begin{equation}
\label{highest}
\epsilon_{max} = 2\times 10^{20}\; eV \eta_2^{1/3} E_{51}^{1/3} n_1^{1/6} 
\theta_{-1}^{-2/3} \xi^{1/2}
\end{equation}
where the ejecta shell has, at the moment of impacting with the interstellar
medium of density $\rho = n_1 m_p \;g\;cm^{-3}$, a Lorenz factor $\eta = 100 
\eta_2$, the shell explosion energy is $E_{51} \times 10^{51}\; erg$ ,
$\theta = 0.1 \theta_{-1}$ the beaming angle, and $\xi$ 
is the magnetic field energy density in units of the equipartition magnetic
field energy density. As mentioned in  paper I, this corrects a small
error of Vietri (1995). When seen from the shell, this 
highest energy proton will have a Lorenz factor $\gamma_p$ given by 
\begin{equation}
\label{gammap}
\gamma_p = 2\times 10^{9} \eta_2^{-2/3} E_{51}^{1/3} n_1^{1/6} 
\theta_{-1}^{-2/3}\xi^{1/2}\;.
\end{equation}
The two equations above correspond to Equations 29 and 28, respectively, of
Vietri (1995).  It is perhaps useful to remind the reader that they were
obtained under the assumption that the major limiting factor of the protons'
energies comes from the finite shell thickness: whenever the proton energy 
increases so much that $g$ times its Larmor radius $r_L$ exceeds the
shell thickness, the proton will cross the whole shell without being 
deflected backwards, and the acceleration cycle stops. Here, 
the magnetic field is assumed to be a fraction $\xi^{1/2}$ of its
equipartition value, and the value $g \approx 40$ is taken from the numerical
simulations of Quenby and Lieu (1989). 

However, at the referee's prompting, I can also show that the acceleration 
time--scale is 
shorter than the shell light--crossing time, which is the time--scale on which 
adiabatic losses set in, and also the time--scale on which the highest energy 
protons cross the shell before being deflected backwards. By establishing this,
I shall in fact demonstrate that the two above equations correctly describe 
the highest proton energies achievable. The acceleration time--scale at
relativistic shocks $t_r$ is shorter by a factor $Q \approx 13.5$ than the
traditional acceleration time--scale at non--relativistic shocks (Quenby and
Lieu 1989); thus 
\begin{equation}
t_r = \frac{t_{nr}}{Q} = \frac{c}{Q(V_1-V_2)} \int_{p_i}^{p_f} \left(
\frac{\lambda_1}{V_1} + \frac{\lambda_2}{V_2}\right)
\frac{d\!p}{p}\;.
\end{equation}
Here matter on the two sides of the shocks moves with speeds $V_1, V_2$
with respect to the stationary shock, 
$\lambda = g r_L$ is the typical deflection length, $p_i$ is the injection
momentum, and $p_f$ the highest achievable momentum in the shock frame. 
Since $\lambda = g r_L = g c p/e B$, I find
\begin{equation}
t_{r} = \frac{c}{Q(V_1-V_2)} \int_{p_i}^{p_f} \frac{g c}{e B}\left(
\frac{1}{V_1}+\frac{1}{V_2}\right) \; d\!p = 
\frac{g c^2 p_f}{(V_1-V_2) Q e B} \left(\frac{1}{V_1}+\frac{1}{V_2}\right)\;.
\end{equation}
Specializing now to a relativistic flow, for which $V_1, V_2 \approx c$, 
and also $V_1-V_2= q c \approx c$, I obtain 
\begin{equation}
t_{r} \approx \frac{2 g p_f}{Q q e B}\;.
\end{equation}
On the other hand, for the highest energy protons, we have set, in deriving 
Eq. \ref{highest}, $\lambda = g r_L =  g c p_f / eB = r_{th}$, with $r_{th}$ 
being the shell thickness; inserting this into the above equation, I find 
\begin{equation}
\label{acctime}
t_{r} = \frac{ 2 r_{th} }{q Q c} \approx 0.1 \frac{r_{th}}{c} \;,
\end{equation}
which is as long as the shell light crossing time. So the acceleration 
is prompt enough to propel protons to the highest energies computed in Eq. 
\ref{highest}, within the time--scale $r_{th}/c$, which is approximately the 
whole time the proton spends inside the shell, before reaching an energy so
large that it cannot be turned backwards. Furthermore, proton adiabatic losses 
also become appreciable on the shell expansion timescale (Rachen and 
M\'esz\'aros 1998) $r_{th}/c$, longer than the acceleration timescale Eq. 
\ref{acctime}. 

It should be noticed that this result is independent of the specific values of 
$g$ assumed; nor does it depend upon the magnetic field having achieved or not 
its equipartition value. It is only the highest proton energy which depends on 
the parameters $g$ and $\xi$. Nor does it depend explicitly upon the shock 
Lorenz factor, provided the shock is relativistic of course. This is especially 
important when one considers the application of this same idea to the bursts' 
afterglows: the maximum proton energy is again provided by the finite shell 
thickness, which however, following Eq. \ref{highest}, changes as the shell 
slows down. The reason for the continuing applicability of Eq. \ref{highest} is 
that, even during the afterglow, the shell thickness is given, in the shell 
frame, by $r_{th} = r /c$ where $r$ is the current shock position, exactly as 
during the burst proper (M\'esz\'aros, Laguna and Rees 1994). This result can be
phrased as follows: as the shock decelerates, the acceleration time--scale
becomes longer, but so does also the shell crossing time and, provided the
magnetic field builds up to the same, constant, fraction of the equipartition
value $\xi$, the two time--scales lengthen by the same amount, thus keeping
the result of Eq. \ref{acctime} still valid. 

The only hidden hypothesis in this argument is that appreciable magnetic fields
exist on both sides of the shock to deflect the cosmic rays; in Vietri (1995)
I suggested that this occurs when two or more shells collide. I believe the
same suggestion is still valid for afterglows; I shall however discuss the
implications of these multiple collisions in a different paper. 

Lastly I should remark that the results discussed in this section do depend
upon the correctness of the assumption made by Quenby and Lieu (1989) that 
scattering of high energy protons at relativistic shocks is isotropic, and 
large--pitch angle. Since this provides a satisfactory explanation 
for the acceleration of cosmic rays at AGNs (Quenby and Lieu 1989), and it is
borne out by theoretical arguments and simulations of interplanetary shocks 
(Moussas, Quenby and Valdes--Galicia 1982, 1987), it seems quite a plausible
assumption to make. 

\section{On fireballs}

The Equations \ref{highest} and \ref{gammap}
are however in an inconvenient form, the reason being 
that the parameter $n_1$ is unobservable. It is expedient to substitute for it
the burst duration $T = T_2\times 100\; s$, a directly observed quantity
which is related to it by the following argument. The shock with the 
interstellar medium forms at a distance $r_{sh}$ from the explosion site, 
where (M\'esz\'aros, Laguna and Rees 1993)
\begin{equation}
r_{sh} = \left(\frac{3 E}{4\pi n_1 m_p c^2 \theta^2\eta^2}\right)^{1/3}\;,
\end{equation}
while the burst duration is of order
\begin{equation}
\label{duration}
T = \frac{r_{sh}}{2\eta^2 c}\;.
\end{equation}
This is an old argument, originally due to Ruderman (1975), and more 
recently discussed by Sari and Piran (1997). Eliminating $r_{sh}$ from the
above I find
\begin{equation}
\label{n1}n_1 = \frac{3 E}{32\pi m_p c^5 \theta^2 \eta^8 T^3} = 8\; cm^{-3}
\; E_{51} \eta_2^{-8} T_2^{-3}\theta_{-1}^{-2}\;,
\end{equation}
and 
\begin{equation}
\label{rshock}
r_{sh} = 2 \eta^2 c T = 6\times 10^{14}\; cm \; T_2 \; \eta_2^2\;.
\end{equation}
The magnetic field behind the shock, in the shell frame, is often assumed
(M\'esz\'aros, Laguna and Rees 1993) to be some fixed fraction $\xi^{1/2}$
of its equipartion value $B_{eq}$, and is given by
\begin{equation}
\label{bequi}
B = (8\pi \xi n_1 m_p c^2)^{1/2}\eta = 54 \; G \; E_{51}^{1/2}
\eta_2^{-3} T_2^{-3/2}\theta_{-1}^{-1} \xi^{1/2}\;, 
\end{equation}
where I have of course used Eq. \ref{n1}. It is also convenient to rewrite 
Eq.\ref{gammap}, again using Eq. \ref{n1}, as
\begin{equation}
\label{gammamax}
\gamma_p = 3\times 10^{9}\; E_{51}^{1/2} \eta_2^{-2} T_2^{-1/2} 
\theta_{-1}^{-1}\xi^{1/2}
\end{equation}
and Eq. \ref{highest} as
\begin{equation}
\label{high2}
\epsilon_{max} = 3\times 10^{20} \; eV E_{51}^{1/2} \eta_2^{-1} T_2^{-1/2} 
\theta_{-1}^{-1} \xi^{1/2}\;.
\end{equation}

\section{Losses}

I have established above that protons suffer no adiabatic losses during
their acceleration; in Vietri (1995) I had shown that also synchrotron
and photopion losses are negligible. In Vietri (1998) I showed that the 
small photopion production leads, through the decays of Eq. 1, to the
production of a small, but significant flux of prompt neutrinos. However,
Rachen and M\'esz\'aros (1998) have shown that one ought to consider 
carefully whether synchrotron and adiabatic losses by muons and pions
before their decay may significantly limit neutrinos' energies. 

The most significant losses will be due to muons, because of their longer 
lifetime ($\tau_\mu = 2.2\times 10^{-6}\; s$, $\tau_\pi = 2.6\times 10^{-8}
\; s$). The maximum muon Lorenz factor $\gamma_\mu$ before significant 
adiabatic losses set in is given by the following argument. A muon moving 
with Lorenz factor $\gamma$ in the shell frame appears to have a lifetime 
$\gamma \tau_\mu$, because of relativistic time dilation. Within this time, 
the shell is expanding, thusly allowing the comoving magnetic field to decrease 
on a time--scale $\approx r_{th}/\dot{r}_{th}$. The shell typical size $r_{th}$ 
in the shell frame is $r/\eta$ (M\'esz\'aros, Laguna and Rees 1993), where 
$r$ is the shock instantaneous distance from the origin of the explosion; 
at the moment of the burst $r = r_{sh}$ (Eq. \ref{rshock}). The shell
expansion velocity $\dot{r}_{th} \approx c$. The limiting Lorenz factor 
$\gamma_\mu$ is found by equating these two time--scales:
\begin{equation}
\label{adiabatic}
\gamma_\mu = \frac{r_{sh}}{\eta c\tau_\mu} = 9\times 10^7 T_2 \eta_2\;.
\end{equation}
The condition that neutrino production is not affected, $\gamma_p < 
\gamma_\mu$, can be expressed as a requirement on the burst parameters: I 
obtain
\begin{equation}
\label{first}
\eta_2 > 3.1 E_{51}^{1/6} T_2^{-1/2} \theta_{-1}^{-1/3}\xi^{1/6} \;.
\end{equation}
For pion adiabatic losses, the only change comes in because of its shorter
lifetime:
\begin{equation}
\label{firstpion}
\eta_2 > 0.5 E_{51}^{1/6} T_2^{-1/2} \theta_{-1}^{-1/3}\xi^{1/6} \;.
\end{equation}

Synchrotron losses by muons are related to those of protons as follows. 
Calling $t_s = 1\; yr\; (10^{11}/\gamma) (1\; G/B)^2$ the lifetime against
synchrotron losses for a proton of Lorenz factor $\gamma$, the largest
muon Lorenz factor $\gamma_\mu$ before significant losses set in is
given by (Rachen and M\'esz\'aros 1998)
\begin{equation}
\gamma t_s \left(\frac{m_\mu}{m_p}\right)^3 = \gamma_\mu^2 \tau_\mu\;.
\end{equation}
Here the muon mass is $m_\mu \approx 0.1 m_p$. I find, using Eq. 
\ref{bequi},
\begin{equation}
\label{synchro}
\gamma_\mu = 4\times 10^{10} \frac{1\; G}{B} = 7.4\times 10^8 E_{51}^{-1/2}
\eta_2^3 T_2^{3/2}\theta_{-1}\xi^{-1/2}\;.
\end{equation}
Again, the condition that these losses are irrelevant, $\gamma_p < 
\gamma_\mu$, can be rewritten as a requirement on the burst parameters: I 
find
\begin{equation}
\label{second}
\eta_2 > 1.3 E_{51}^{1/5} T_2^{-2/5}\theta_{-1}^{-2/5}\xi^{1/5}\;. 
\end{equation}
For pions the relevant criterion is
\begin{equation}
\label{secondpion}
\eta_2 > 0.4 E_{51}^{1/5} T_2^{-2/5}\theta_{-1}^{-2/5}\xi^{1/5}\;. 
\end{equation}

It is convenient to remark here that the use of $\eta$ and $T$ as independent
burst parameters allows discussion of afterglow with the same formulae. 
During afterglows, the quantity $T$ loses meaning because the afterglow emission
is of course continuous. However, let us still define, by analogy with Eq. 
\ref{duration}, $T \equiv \frac{r}{2\eta^2 c}$, 
where $r$ and $\eta$ are the instantaneous shock position and Lorenz factor.
During the afterglow, for adiabatic expansion the shell Lorenz factor
decreases as $\eta \propto r^{-3/2}$. Thus, the variation of $T$ parametrizes 
the shell radius. I find that
\begin{equation}
\label{afterglow}
\eta = \eta_i \left(\frac{T}{T_i}\right)^{-3/8}
\end{equation}
describes the further evolution of the shell. Here $\eta_i$ and $T_i$ are 
the initial values of the afterglow, at the moment of the formation of
the external shock. This equation can also be established, for an 
adiabatic expansion in a constant density environment,by imposing the 
constancy through the afterglow of Eq. \ref{n1}. It is also identical to the 
time evolution law for adiabatic expansion with respect to observer's time, 
so that the fictitious quantity $T$ can also be identified with Earth time, 
a physical quantity. In other words, by considering arbitrary values of
$\eta$ and $T$, I am considering points which do not really model any
known burst, but which parametrize correctly later, afterglow moments of
realistic bursts; the realistic initial models are linked to later 
values of $\eta$ and $T$ by Eq. \ref{afterglow}. 

First I consider neutrinos produced directly by charged pion decay. In 
Fig. 1 I show the constraints Eq. \ref{firstpion} and \ref{secondpion}
for $E_{51} = \theta_{-1} = \xi=1$, plotted as solid lines. 
The dashed lines represent the afterglow evolution tracks, Eq. \ref{afterglow}. 
The figure covers roughly the first three afterglow hours. The dotted lines 
represent the loci of points of constant highest neutrino energies: when
losses are negligible, these are computed multiplying highestproton energies,
Eq. \ref{high2}, by $0.05$. 
Neutrinos produced directly by charged pion decay will not be limited by the 
rather more stringent requirements on muon losses, Eq. \ref{first}. 
This means that, should a burst fail to produce high energy neutrinos from 
muon decay, its neutrino flux will only be decreased (roughly) by a factor 
of 3 because the
two neutrinos from the muon decay will be lost, at the same energy level.
It can easily be seen from Fig. 1 that neutrinos of energy as large as 
$10^{19}\; eV$ may be produced. 

Muon--produced neutrinos will be limited in energy both by synchrotron 
losses and by adiabatic losses, depending exactly on the model parameter 
values. The cross--over occurs when Eq. \ref{adiabatic} equals Eq. 
\ref{synchro}, \ie,for
\begin{equation}
\label{crossover}
\eta_{2cr} = 0.3 E_{51}^{1/4} T_2^{-1/4} \theta_{-1}^{-1/2}\xi^{1/4}\;
\end{equation}
with adiabatic losses dominating whenever $\eta_2 > \eta_{2cr}$. 

When synchrotron losses dominate we see by comparing Eq. \ref{synchro} with 
Eq. \ref{afterglow} that afterglow evolution occurs along lines of (roughly !)
constant $\gamma_\mu$ in the shell frame, so that the maximum achievable 
energy in the observer frame is given by $100\times \eta_2$ times Eq. 
\ref{synchro}. Multiplying times the typical energy of outgoing neutrinos, 
$\approx 50\; MeV$ we find that the neutrino energy is
\begin{equation}
\label{synchro2}
\epsilon_\nu^{(sy)} = 3.7\times10^{18}\; eV E_{51}^{-1/2} \eta_2^4 T_2^{3/2}
\theta_{-1}\xi^{-1/2}\;. 
\end{equation}
Taking into account that during afterglow the shell Lorenz factor varies 
according to Eq. \ref{afterglow}, the typical neutrino energy is given by
inserting this into the above equation:
\begin{equation}
\label{emax1}
\epsilon_\nu = 3.7\times 10^{18} \; eV \eta_{21}^4 T_{2i}^{3/2} 
\xi^{-1/2}
\end{equation}
where of course $\eta_{2i}$ and $T_{2i}$ are the initial Lorenz factor
and burst duration in units of $100$ and $100\; s$, respectively, and the
result is independent of post--burst time.  Since the dependence on the 
initial parameter is so steep, it is enough that 
$\eta_{2i}$ exceeds unity by a factor of $2$  for $\epsilon_\nu$ to exceed
$10^{19}\; eV$. When adiabatic losses dominate, the limiting muon Lorenz 
factor in the observer frame is given by $\gamma_\mu^{(obs)} = 9\times 
10^9\; T_2 \eta_2^2$. The locus of constant $\gamma_\mu^{(obs)}$ is thus 
parallel to the top solid line of Fig. 1 (Eq. \ref{first}), and to Eq. 
\ref{high2}. This gives a typical neutrino energy of
\begin{equation}
\label{adiabatic2}
\epsilon_\nu^{(ad)} = 4.5\times 10^{17}\; eV T_2 \eta_2^2\;.
\end{equation}
However, afterglow evolution (Eq. \ref{afterglow}) is shallower than either 
of these; thus during the afterglow the energy of neutrinos produced becomes 
higher; this can be checked by computing the highest $\gamma_\mu^{(obs)}$ 
for the afterglow evolution, Eq. \ref{afterglow}: I find $\gamma_\mu^{(obs)} 
\propto T^{1/4}$. The normalization is such that the highest neutrino energy 
occurs on the top solid line of Fig. 1: inserting Eq. \ref{first} into Eq. 
\ref{high2}, and considering that typical neutrinos carry away $0.05$ of the 
proton energy I find
\begin{equation}
\label{emax2}
\epsilon_\nu = 5\times 10^{18}\; eV\; E_{51}^{1/3} \theta_{-1}^{-2/3}
\xi^{1/3}\;.
\end{equation}

In Fig. 2 I show as dashed lines tracks of constant energy for neutrinos 
produced by muon decay only, in the observer frame, together with the previously
defined afterglow tracks (Eq. \ref{afterglow}, again dashed lines). The bottom
solid line, Eq. \ref{crossover}, marks the boundary between the region where 
adiabatic losses (above) and that where synchrotron losses dominate (below). 
The top solid line marks the region (below the line) where adiabatic muon 
losses limit neutrinos' energies. 
At fixed $T_2$, the neutrino energy is not monotonic
with $\eta_2$ because, above the region where adiabatic losses operate,
each neutrino carries away a fraction $\approx 0.05$ of the emitting proton 
energy, which is a decreasing function of $\eta_2$, for fixed $n_1$ (Eq.
\ref{high2}). In either case, \ie, synchrotron or adiabatic losses, we see that
it is still possible to produce high energy neutrinos, of order $10^{19}\; eV$.

In paper I, I showed that the fraction $f_\pi$ of the photon luminosity 
$L_\gamma$ that goes into neutrinos is
\begin{equation}
\label{fpi}
f_\pi = 0.01 \eta_2^{-4}\frac{300\; keV}{\epsilon_\gamma} T_2^{-1}
\end{equation}
where $\epsilon_\gamma \approx 300-10^3\; keV$ is the typical burst 
observed spectral turnover energy. The important parameter $\epsilon_\gamma$
and its time--dependence during the afterglow are not currently observed, 
nor are they in any way predicted by theory. A simple argument allowed to
state that, if $\epsilon_\gamma$ scales with the shell Lorenz factor $\eta$ as 
$\propto \eta^a$ with $a\geq 1$, then $f_\pi$ remains constant or grows through 
the afterglow; if $f_\pi$ remains constant, then the 
neutrino luminosity $L_\nu = f_\pi L_\gamma \propto t^{-1}$, the last
proportionality being observed (Fruchter \etal, 1997). Thus, even the bursts
which, initially, suffer important muon losses, during the afterglow 
become effective neutrino emitters and, given that the neutrino luminosity
scales only with $t^{-1}$, the total neutrino fluence will not be much 
reduced by losses. The reason why I find now that, initially, losses are 
important is that,
when I assume a significant amount of beaming, the maximum proton Lorenz factor,
Eq. \ref{gammamax}, is greatly increased, while the maximum muon Lorenz factor,
Eq. \ref{adiabatic} for adiabatic losses, or Eq. \ref{synchro} for 
synchrotron losses, is reduced or at most remains constant; it can be seen that 
in the isotropic case assumed in paper I the requirement on losses would have
been less severe; however, as a partial compensation, for same total energy but
significant beaming the maximum proton energy in the observer frame also 
increases, so that the end result is that, losses or not, neutrinos with 
energies exceeding $10^{18}\; eV$ are emitted. 

\section{Discussion}

The propagation of ultra high energy cosmic rays in intergalactic space is
significantly limited by energy degradation via pion photoproduction off
the CMBR photons (Greisen 1966, Zatsepin and Kuz'min 1966). However, it has
long been recognized (Wdowczyk, Tkaczyk and Wolfendale 1972) that observations
of high energy neutrinos, thanks to their negligible cross sections for 
interaction with matter and photons, will eventually allow us to investigate
the sources of cosmic rays way beyond this limit. With the development of
Airwatch--class experiments (Linsley 1997), capable of monitoring from space
atmospheric areas of order $10^6\; km^2$ in the search for extended air showers,
and with detection efficiencies close to $1$ for neutrinos (L. Scarsi, private
communication), it seems the time for extending our search for sources of ultra
high energy cosmic rays to the whole Universe has come.

Two comments are in order. First, as already noted in paper I, the generic 
neutrinos (\ie, those produced by cosmic rays which managed to escape unscathed
from their sources, but which produce pions off CMBR photons) should display
dipole and quadrupole moments which, if UHENs come from GRBs, should be totally 
negligible (Fishman and Meegan 1995). This by itself may help rule out/establish
an important alternative idea, that UHECRs have an origin in the Local 
Supercluster (Stanev \etal, 1995). Second, the expected neutrino fluxes computed
by Yoshida and Teshima (1993) for cosmologically distributed sources, leading to
expected event rates for Airwatch--class experiments of $200 \; yr^{-1}$, may 
be a significant underestimate. In fact, they assumed a distribution of sources
of UHECRs arbitrarily limited to a redshift $z_{max} = 2$. However, the recent,
sensational identification of a redshift $z = 3.43$ for the burst GRB 971214
(Kulkarni \etal, 1998) makes it clear that significant fluxes of ultra high
energy neutrinos may be at very large redshifts, thus not violating the
constraints on fluxes of UHECRs observed at Earth. Thus the building of 
Airwatch--type experiments with significant neutrino--detection capabilities 
becomes an even more exciting prospect. 

In this paper, I have strengthened the point made in paper I that a fraction
$f_\pi \approx 0.01$ of all ultra high energy neutrinos should correlate
(within about a week) with the burst where it was emitted, by showing in some
detail that losses do not inhibit the production of high energy ($\ga 
10^{19}\; eV$) neutrinos, simultaneously with the burst or its afterglow.
A search for ultra high energy neutrinos ought to be a significant test of
the hypothesis that UHECRs are generated in GRBs.

It is a pleasure to acknowledge helpful comments from J. Rachen and P. 
M\'esz\'aros, and from an anonymous referee.

\newpage

\begin{figure}
\centerline{\psfig{figure=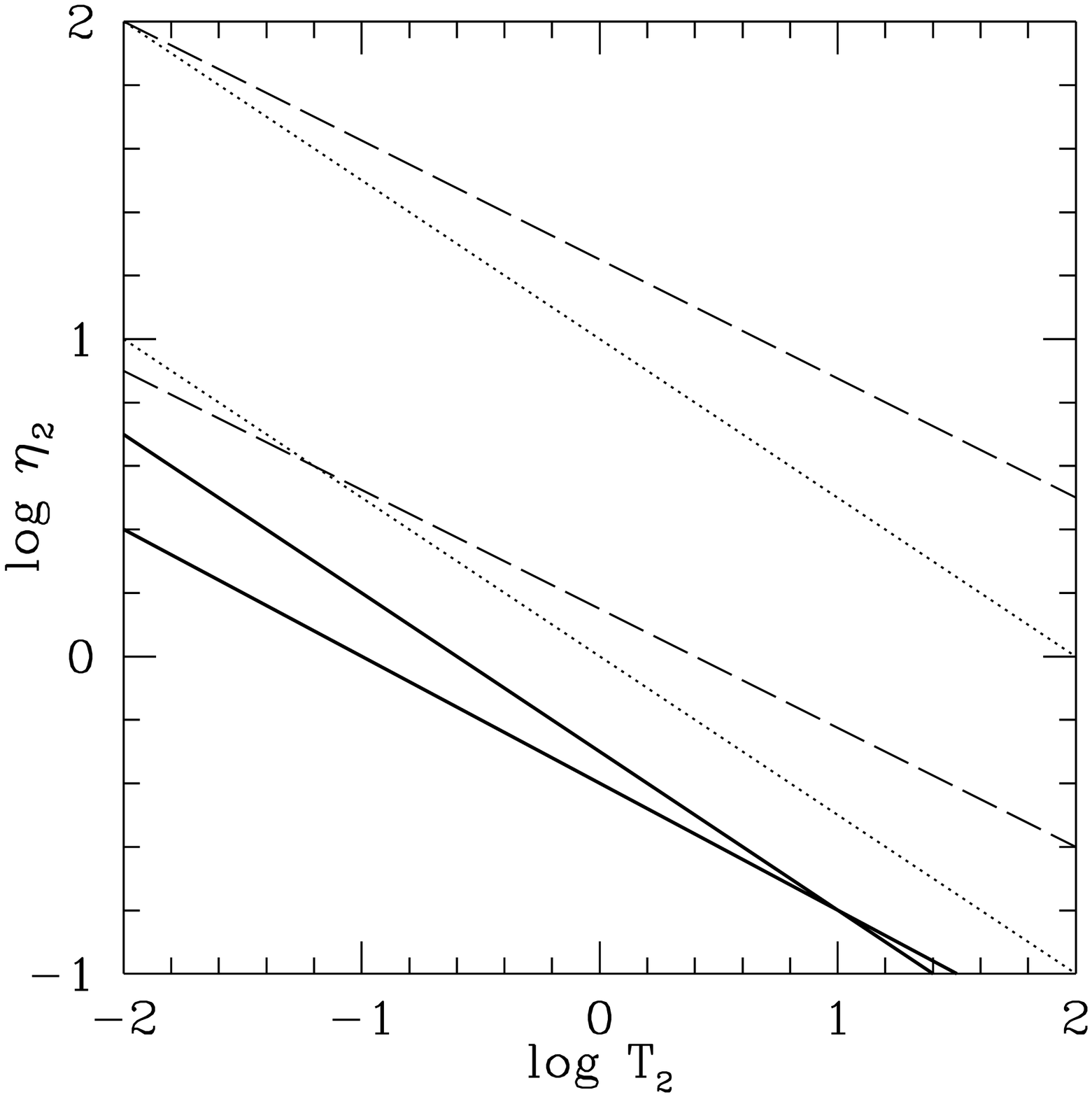,width=11cm,height=15cm}}
\caption[]{Typical energies for neutrinos produced by charged pion decay,
as a function of bursts' parameters. 
Top solid line (Eq. \ref{firstpion}): pion adiabatic losses
are negligible in the upper region. Bottom solid line: Eq. \ref{secondpion}.
Below this line synchrotron losses become important. Dashed lines: 
representative tracks for afterglow evolution, Eq. \ref{afterglow}. Dotted 
lines: loci of points of constant neutrino energies in the 
observer's frame. The top dotted line corresponds to $\epsilon_\nu = 
1.5\times 10^{18}\; eV$, while the bottom dotted line to $\epsilon_\nu = 
1.5\times 10^{19}\; eV$. These values are obtained by multiplying Eq. 
\ref{high2}  times $0.05$, the typical yield for the neutrino energy. 
The figure covers, roughly, the first three afterglow hours.}
\label{Figure 1}
\end{figure}

\newpage

\begin{figure}
\centerline{\psfig{figure=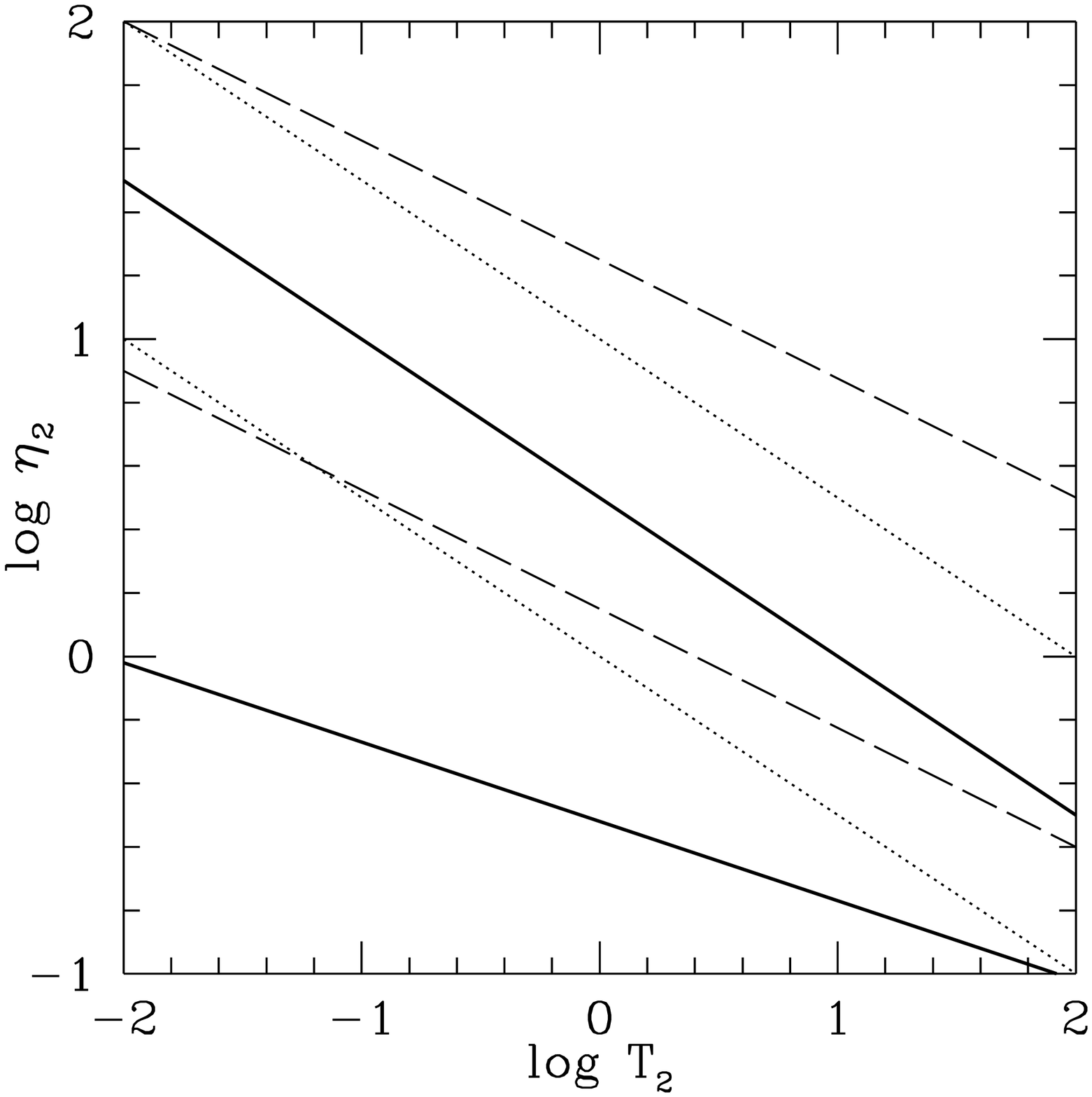,width=11cm,height=15cm}}
\caption[]{Typical energies for neutrinos produced by muon decay as a 
function of bursts'parameters. The bottom solid line (Eq. \ref{crossover}) 
separates the region where adiabatic losses prevail (above the line) over 
synchrotron losses. The top solid line solid line (Eq. \ref{first}) 
separates regions where adiabatic losses are important (below the line)
from that where where they are negligible (above the line). 
Dotted lines: loci of constant neutrino energy, in the 
observer's frame. The top dotted line corresponds to $\epsilon_\nu = 
1.5\times 10^{18}\; eV$, while the bottom dotted line to $\epsilon_\nu = 
4.5\times 10^{17}\; eV$. Dashed lines: representative afterglow tracks, Eq. 
\ref{afterglow}. Since afterglow tracks are shallower than constant neutrino 
energy tracks, it can be seen that later in the afterglow emitted neutrinos 
will have larger energies.}
\label{Figure 2}
\end{figure}


\begin{references}
\reference{} Costa, E., \etal, 1997, Nature, 387, 783.
\reference{} Eichler, D., Livio, M., Piran, T., Schramm, D.N., 1989, Nature, 
340, 126. 
\reference{} Fishman, G.J., Meegan, C.A., 1995, \araa, 33, 415.
\reference{} Frail, D.A., \etal, 1997, Nature, 389, 261.
\reference{} Fruchter, A., \etal, 1997, IAU Circ. n. 6747.
\reference{} Goodman, J., 1997, New Astronomy, 2, 449. 
\reference{} Greisen, K., 1966, \prl, 16, 748.
\reference{} Heise, J., \etal, 1997, IAU Circ. n. 6787.
\reference{} Janka, H.-T., Ruffert, M., 1996, \aap, 307, L33.
\reference{} Kulkarni, S., \etal, 1998, Nature, 393, 35. 
\reference{} Linsley, J., 1997, in {\it Proc. of the XXVth ICRC
Conference}, Durban, RSA, 5, 381.
\reference{} M\'esz\'aros, P., Laguna, P., Rees, M.J., 1993, \apj, 405, 278.
\reference{} Moussas, X., Quenby, J.J., Valdes--Galicia, J.F., 1982, ApSS, 
86, 185.
\reference{} Moussas, X., Quenby, J.J., Valdes--Galicia, J.F., 1987, Solar 
Phys., 112, 365. 
\reference{} Narayan, R. Paczynski, B., Piran, T., 1992, \apjl, 395, L83.
\reference{} Ostrowski, M., Zdiarski, A.A., 1995, ApSS, 231, 339.
\reference{} Paczynski, B., Xu, G., 1994, \apj, 427, 708. 
\reference{} Panaitescu, A., M\'esz\'aros, P., Rees, M.J., 1998, \apj, in press,
astro-ph. n. 9801258.
\reference{} Quenby, J.J., Lieu, R., 1989, Nature, 342, 654.
\reference{} Rachen, J., M\'esz\'aros, P., 1998, submitted to Phys. Rev. D,
astro-ph. n. 9802280.
\reference{} Rees, M.J., M\'esz\'aros, P., 1992, \mnras, 258, 41P. 
\reference{} Ruderman, M., 1975, Ann. N.Y. Acad. Sci., 262, 164.
\reference{} Sari, R., Piran, T., 1997, \apj, 485, 270.
\reference{} Stanev, T., Biermann, P.L., Lloyd-Evans, J., Rachen, J. Watson, 
A.A., 1995, \prl, 75, 3056.
\reference{} van Paradijs, J., \etal, 1997, Nature, 386, 686.
\reference{} Vietri, M., 1995, \apj, 453, 883.
\reference{} Vietri, M., \prl, 80, 3690.
\reference{} Waxman, E., 1995, \prl, 75, 386.
\reference{} Waxman, E., Bahcall, J., 1997, \prl, 78, 2292,
\reference{} Wdowczyk, J., Tkaczyk, W., Wolfendale, A.W., 1972, J. Phys. A, 5,
1419.
\reference{} Zatsepin, G.T., Kuz'min, V.A., 1966, JETP Lett., 4, 78. 
\end{references}
\end{document}